generoso.immediato@gmail.com

# Epistemic Trade-Off: An Analysis of the Operational Breakdown and Ontological Limits of "Certainty–Scope" in AI

*Generoso Immediato*

generoso.immediato@gmail.com

## *Abstract*

*The recently published "certainty-scope" conjecture offers a compelling insight into the inherent trade-off present within artificial intelligence (AI) systems. As general research, this investigation remains vital as a philosophical undertaking and a potential guide for directing AI investments, design, and deployment, especially in safety-critical and mission-critical domains where risk levels are substantially elevated. While maintaining intellectual coherence, its formalization ultimately consolidates this insight into a suspended epistemic truth, which resists operational implementation within practical systems. This paper argues that the conjecture's objective to furnish insights for engineering design and regulatory decision-making is limited by two fundamental factors: first, its dependence on incomputable constructs and its failure to capture the generality factors of AI, rendering it practically unimplementable and unverifiable; second, its foundational ontological assumption of AI systems as self-contained epistemic entities, distancing it from the complex and dynamic socio-technical environments where knowledge is co-constructed. We conclude that this dual breakdown—an epistemic closure deficit and an embeddedness bypass—hinders the conjecture's transition to a practical and actionable framework suitable for informing and guiding AI deployments. In response, we point towards a possible framing of the epistemic challenge, emphasizing the inherent epistemic burdens of AI within complex human-centric domains.*

generoso.immediato@gmail.com

## Introduction

Unsurprisingly, the epistemic limits of artificial intelligence (hereinafter referred to as AI) are among the most relevant concerns in its design, deployment, and governance; notably, identifying and addressing them is currently on the agenda of both the academic and business worlds. As AI systems expand across (i) *operational domains* and (ii) *levels of autonomy*, a critical question[1] arises: *how can we jointly provide requirements for generality and certainty, and which trade-offs constrain that joint satisfaction, impact performance, and lifecycle cost?*

The philosopher Luciano Floridi, in response to these timely and challenging questions, proposes a formal conjecture: a fundamental trade-off between AI's certainty and scope (Floridi, 2025a; Floridi, 2025b). Grounded in philosophical reflections across AI's historical evolution, the conjecture posits a persuasive and intuitive claim, as he writes, «*as certainty about AI's outcomes increases, scope must shrink, that is, as AI systems' scope grows to cover complex, unstructured domains and corresponding powerful applications, perfect certainty about the reliability of their results becomes unattainable. No system can escape this trade-off*».

While fully recognizing and valuing the intellectual spark that inspired this formulation, we argue that, as published, the conjecture resists operationalization and risks collapsing under its own abstraction, as it does not sufficiently account for the *epistemic entanglement*[2] of hybrid[3] AI within the complex reality of socio-technical systems.

---

[1] This is, in fact, a triple question: (i) conditional on the declared variables, there may be a clear *implementation-level* relationship; (ii) there is a relationship with *admissible performance*, under specified technical constraints and within a given application domain and usage context; and (iii) there may be a *competitiveness/cost trade-off.* It must be stressed that these dimensions are distinct and depend on the initial assumptions—particularly on whether certainty is construed as (a) an *intrinsic feature of the AI system*, or (b) a *measure of implementation error*, i.e., deviation from the specified output or expected performance. In our view, these distinctions are important: there is no universal answer, and the analysis must be anchored to clearly stated assumptions.

[2] In this context, *epistemic entanglement* denotes, in practice, the non-isolation of AI systems: generally speaking, despite being an *epistemic technology* (Alvarado, 2023), *the epistemic content* AI produces necessarily depends on interactions with the *operational environment* and *human oversight.*

[3] **Hybrid** (as used here) denotes a heterogeneous AI system in which symbolic (e.g., logic, knowledge graphs, constraints) and subsymbolic (e.g., learned embeddings, neural networks) representations and procedures co-exist and are orchestrated within a single system boundary. Hybridity spans: (i) *representational hybridity* (discrete, interpretable structures alongside continuous, learned representations); (ii) *inferential hybridity* (joint use of data-driven learning with deductive/constraint-based reasoning, including mechanisms such as constraints-as-losses and programmatic priors); and (iii) *operational hybridity* (runtime arbitration, calibrated abstention, and human oversight within a declared Operational Design Domain (ODD)—the set of operating conditions and constraints under which the system is designed to function safely; cf. SAE J3016, 2021). Accordingly, hybrid is not a midpoint between symbolic and subsymbolic approaches but a system-level property of integration and coordination across the socio-technical arrangement over the lifecycle (cf. d'Avila Garcez & Lamb, 2009; De Raedt et al., 2018; SAE J3016, 2021).

generoso.immediato@gmail.com

# 1. Deconstructing the Conjecture

The reasoning behind the conjecture can be depicted in four steps:

1. **Observation**: AI systems face a tension between provable certainty and expressive generality.
2. **Philosophical Framing**: This tension reflects a deeper epistemic trade-off, with epistemological implications for how we model and govern AI.
3. **Formalization**: An inequality[4] involving certainty and scope, linking the latter to Kolmogorov complexity[5].
4. **Lack of Operational Closure**: Absent a computable, auditable instantiation, the conjecture lacks operational closure and therefore falls short as a design or governance metric.

The constraints begin to emerge in Step 3, for it inherits the ontological assumptions—summarized below—on which the epistemological framing rests.

---

[4] ***Corrigendum* Note (1 October 2025):** the originally published inequality $1 - C(M) \cdot S(M) \geq k$ was misstated (Floridi, 2025a); the author has corrected it to $C(M) \cdot S(M) \leq k$, with $k \in (0, 1)$, and aligned the corollaries/figure accordingly (Floridi, 2025b). For transparency and rigor, we reflect the corrected form; this amendment is immaterial to the remainder of our analysis.

[5] **Kolmogorov complexity** – '$\mathrm{K}(\cdot)$' (Li & Vitányi, 2019), formally defined as the length of the shortest binary program that outputs a given string, is a canonical example of an incomputable function—no general algorithm exists that can compute it for arbitrary inputs (this follows from the *halting problem* and related undecidability results (Turing, 1937; Rice, 1953)). While, for example, *compression-based* approximations are used in practice, these offer upper bounds on Kolmogorov complexity and are implementation-dependent. This limitation poses a significant challenge for engineering practice: any metric built upon Kolmogorov complexity inherits its non-verifiability, making it unsuitable for domains requiring auditable, testable, and bounded epistemic measures—especially in the field of safety-critical systems. Moreover, such abstraction may inadvertently obscure compatibility with established inference frameworks—such as *Bayes-optimal reasoning* (which, under explicit model assumptions, provides computable mechanisms). In this light, while the conjecture's formulation captures an intuitive trade-off, its reliance on incomputable constructs makes it difficult to reconcile with the methodological standards already established in AI validation, scientific modeling, and decision science. ***Remark***: For the avoidance of doubt, our critique pertains *solely* to the way Kolmogorov complexity is employed within the conjecture; we do not question its theoretical robustness or its established relevance in appropriate contexts. The central issue lies in the scope definition: $\mathrm{S(M)} \coloneqq \mathrm{K(I)} + \mathrm{K(O)}$, where $\mathrm{I}$ and $\mathrm{O}$ are the input/output sets (Floridi, 2025a; Floridi, 2025b). This formulation measures only the compressibility of the input/output representations (Shannon, 1948; Weaver, 1949). It does not account for (i) the internal algorithmic implementation and representational scheme of the AI mechanism, or (ii) the semantics of knowledge co-constructed with humans and the environment (Weaver, 1949; Wittgenstein, 1953; Suchman, 1987; Jasanoff, 2004). In our view, this results in an ill-posed formalization of the original claim: semantic gap apart, a scope mapping should capture task and function coverage—and thus *generality*—within a given AI mechanism, rather than merely the compressibility of its $\mathrm{I/O}$ sets.

generoso.immediato@gmail.com

- **Boundary assumption**: AI systems are framed in a "*symbolic vs subsymbolic*" dichotomy, with each class occupying distinct poles of the "certainty–scope" spectrum. This framing implicitly defines the two extremes of the conjecture.
- **Ontological assumption**: AI systems are treated as self-contained epistemic entities—abstract generators of scope and certainty holders—without accounting for their entanglement with the socio-technical environments in which they operate (Jackson, 2019; Vermaas et al., 2011).

This raises immediate key questions: (i) *in an era of increasingly hybrid and socio-technically embedded AI systems, can we still uphold such clear-cut boundaries?* (ii) *More importantly, can we assume that epistemic measures like "certainty" and "scope" are intrinsic to the AI mechanism, or are they co-constructed through domain variability, human oversight*[6]*, and system context?* These questions are supported by decades of evidence that cognition and knowledge are distributed across human–artifact systems, not isolated in the artifact (Suchman, 1987; Hutchins, 1995; Jasanoff, 2004).

Inherently, the published conjecture cannot presently be tested or applied in a verifiable way—nor can it be framed for proof or falsification[7] (Popper, 1962). Along these lines, a *Peircean*, fallibilist framework requires that any AI-relevant trade-off be rendered computable, auditable, and falsifiable—as a measurable reduction of *residual doubt*—rather than posited as a universal relation over incomputable quantities that preclude operational closure (Peirce, 1992). As a result, the transition from Step 3 to Step 4 leaves the *observed problem* without a *measurable insight* that links it to its underlying structure. This leads to another critical question: *how can scientists, engineers, and managers rely on an uncomputable inequality as a meaningful measure of epistemic trade-offs in systems they must design, validate, and govern in deployed contexts?*

Furthermore, because the conjecture's primitives are cast as static, the proposed trade-off neglects the temporal and organizational coupling observed in distributed work systems (Suchman, 1987; Hutchins, 1995). Consequently, it cannot inform real-world AI design, implementation, and governance, which are inherently time-varying and iterative.

---

[6] **Human oversight** is used here in the formal sense defined by the author's manifesto (Immediato, 2025): a measurable construct capturing the total human function required to supervise AI outputs and their adoption, specified as *Human Oversight Effort = Cognitive Verification + Ethical Oversight*. **Cognitive Verification** is an *epistemic validation* activity in which the human assesses the meaning, trustworthiness, and applicability of a machine-generated output, rather than verifying the entire system. **Ethical Oversight** is a dedicated human function that oversees the ethical dimensions of an AI system across its lifecycle—from early design onward through development, deployment, monitoring, and decommissioning.

[7] The present analysis prompts inquiries regarding testability and applicability; however, the ambiguous and informal characteristics of the conjecture primitives (Floridi, 2025b) render logical–mathematical proof or disproof unattainable unless the formal framework is meticulously re-evaluated, whilst maintaining the original published primitives unaltered.

generoso.immediato@gmail.com

## 2. Reframing the AI Epistemic Challenge

What matters is not whether AI's generality reduces certainty in principle, but whether we can (i) model, (ii) measure, and (iii) manage that tension within deployed complex hybrid systems.

This requires moving beyond theoretical complexity toward frameworks that are:

- Grounded in **computable functions or heuristics capable of representing time-variant epistemic behaviors**.
- Structured within **bounded models** that enable verification and validation.

As outlined in Section 10 of the *Philosophy of the Machines* (Immediato, 2025), such models must account for the limits imposed by:

- **Computational machinery and algorithms**, characterized by their theoretical performance (*foresight utility*) shaped through domain-specific constraints ($\Delta \cdot F$).
- **Human oversight** captured as epistemic influence ($\eta \cdot H$).
- **System friction and contextual variability**, introducing uncertainty and resistance ($\zeta$).

This is the missing link in the conjecture formulation: an epistemic model capable of returning to the source problem—a real-world *design and deployment* challenge—and offering not only theoretical clarity but operational insight. Factually, the conjecture models AI systems as mechanisms that bear an *epistemic tension* between certainty and scope but are not epistemic agents themselves. Although machines occupy a structural role—modeled, constrained, and evaluated through epistemic lenses—they do not participate in knowledge production as autonomous subjects. Even though Floridi acknowledges the broader complexity of AI deployment, the conjectured trade-off is formalized independently of the system's operational context: it abstracts away the socio-technical entanglement—human oversight, organizational procedures, and environmental constraints that often determine an AI system's epistemic profile. This conceptually resonates with the systems-theoretic intuition expressed by Ashby's *Law of Requisite Variety*[8]: reliable performance in open environments typically requires either (i) *reducing environmental variety* or (ii) *increasing regulator variety* (Ashby, 1956); in other words, if an AI is abstracted as a self-contained epistemic entity, it will inevitably face variety it cannot absorb. The epistemic relevance of AI systems does not stem from their knowledge possession (Ferrario, 2024), but rather in their contribution —under human oversight and within domain constraints—to structured knowledge generation, verification, and utilization processes. Properties such as epistemic certainty and scope, therefore, cannot be meaningfully assigned without reference to this broader system of co-construction.

[8] Wiener (1948, 1950) frames feedback and communication for regulation, while Ashby (1956) formalizes its *variety requirement*—often summarized as "*only variety can destroy variety.*"

The theoretical trade-off does not fail as intuition but collapses as a usable construct — it initiates a valid philosophical query, yet withholds the operational framing required to resolve it within the systems where it seeks relevance. This limitation becomes particularly acute in the context of mission-critical and safety-critical applications. In such domains, any metric—deterministic or heuristic—must be grounded in measurable constructs and domain-specific assumptions; in high-stakes decisions, this often implies preferring interpretable models over post-hoc explanations of black boxes, where accurate interpretable models are available (Rudin, 2019). Absent such grounding, it risks generating epistemically misaligned signals: metrics that appear principled, yet fail to map onto the scale, risk profile, or investment constraints of real-world environments. In this light, heuristics are not objectionable per se — but to serve their function, they must rest on transparent hypotheses and traceable epistemic variables. Only then can such metrics be meaningfully integrated into design-validation workflows and early-stage investment gates[9].

## 3. Conclusion

The "certainty-scope" conjecture stands as an insightful philosophical contribution; however, its formalization on incomputable quantities—incapable of fully encapsulating the essence of generality, breadth, and semantic properties in AI systems, specifically in the definition of scope—precludes a computable and auditable instantiation, thereby constraining its practical utility. Whether or not one accepts a "certainty–scope" tension in practice, efforts to achieve epistemic closure through such published primitives leave unresolved the operational translation of the "certainty" and "scope" abstractions. We do not reject the existence of tension in principle; we argue for the need for an appropriate, computable formalization. Any meaningful AI trade-off framework must achieve operational closure at the system level—or *epistemic entanglement*—which is crucial for incorporating human oversight and system interactions while addressing the time-varying nature of socio-technical environments.

Incomputability is not inherently flawed until it is misconstrued as an operational constraint. At that juncture, it transitions from a theoretical boundary to an epistemic liability. Absent operational closure, any purported universal relation risks devolving into a context-dependent empirical heuristic, if proven at least. From a system engineering and management angle, this prompts a pertinent inquiry: *To what degree is it justifiable*

[9] In practice, *testing, evaluation, verification and validation* (TEVV) is embedded across the lifecycle and can begin at design (NIST, 2023). Early-phase trade studies and gated decisions determine cost and investment (NASA, 2017; Cooper, 2017). ISO/IEC 23894 emphasizes integrating AI risk management into organizational activities—i.e., operational governance alignment (ISO/IEC, 2023).

generoso.immediato@gmail.com

*to devote resources to assessing a heuristic constraint, and what tangible benefits—if any—can such endeavors feasibly yield in real-world systems?*

From our perspective, this question pertaining to AI cannot be resolved in theoretical limits only; it demands evaluation along the foundational dimensions of quality, time, and cost—the core pillars of any engineering, product, and governance strategy. Ultimately, the fundamental epistemic tension persists as both urgent and unresolved; the current analysis invites further interdisciplinary collaboration in forging practical frameworks for complex systems, especially those in AI that engender significant interest and risks.

## Acknowledgements

The author wishes to thank Antonella Migliardi for her invaluable assistance and curatorial support in developing this paper. Her role as curator has been essential to both the conceptual refinement and the formal presentation of the argument.

generoso.immediato@gmail.com

# Table of Contents

---